\documentstyle[preprint,prl,aps]{revtex}
\begin{document}
\draft
\title{Excitation Spectrum and Collective
Modes of Composite Fermions}
\author{X.G. Wu and J.K. Jain}
\address{Department of Physics, State University of New York
at Stony Brook, Stony Brook, New York 11794-3800}
\date{March 15, 1994}
\maketitle
\begin{abstract}

According to the composite fermion theory, the interacting electron
system at filling factor $\nu$ is equivalent to the non-interacting
composite fermion system at $\nu^*=\nu/(1-2m\nu)$, which in turn is
related to the non-interacting electron system at $\nu^*$.
We show that several eigenstates of non-interacting electrons
at $\nu^*$ do not have any partners for interacting electrons
at $\nu$, but, upon composite fermion transformation,
these states are eliminated, and the remaining states provide a good
description of the spectrum at $\nu$. We also show that the
collective mode branches of incompressible states are
well described as the collective modes of composite fermions.
Our results suggest that, at small wave vectors,
there is a single well defined
collective mode for all fractional quantum Hall states.
Implications for the Chern-Simons treatment of
composite fermions will be discussed.

\end{abstract}

%\narrowline

{\bf I. INTRODUCTION}

Several spectacular phenomena have been observed in two-dimensional
electron-systems (2DES) under the
application of a strong transverse magnetic field ($B$). These are
consequences of the formation of a new kind of particle in this system,
called composite fermion (CF), which is an electron
carrying an even number of vortices of the wave function \cite {Jain}.
Composite fermions are formed because, in a certain range of filling
factors, electrons avoid each other most efficiently by capturing
an even number of vortices of the wave function and transforming into
composite fermions. The residual interaction between composite fermions
is weak, and they can be treated as non-interacting to a good first
approximation. The fundamental property of composite fermions
is that they experience an effective field $B^*=B-2m\phi_{0}\rho$,
where $\rho$ is the 2D density of electrons, and $\phi_{0}=hc/e$
is the quantum of flux. Thus, the liquid of strongly correlated
electrons at $B$ is equivalent to a gas of weakly interacting
composite fermions at $B^*$.

The wave functions of non-interacting composite fermions are
constructed as follows \cite {Jain}. Consider non-interacting electrons at
filling factor $\nu^*$, and denote their eigenstates by
$\Phi_{\nu^*,n,\alpha}$. (The filling factor is defined as
$\nu=\rho\phi_{0}/B$.) The spectrum of
non-interacting electrons contains bands of degenerate states
separated by the cyclotron energy $\hbar\omega_{c}$; $n$
is the band index, and $\alpha$ labels the eigenstates within a band.
To write the wave functions of {\em non-interacting} composite
fermions, we attach $2m$ vortices to each electron by multiplying
$\Phi$ by $D^m$, where
the Jastrow factor
\begin{equation}
D\equiv \prod_{j<k}(z_{j}-z_{k})^{2}\;\;,
\end{equation}
$z_{j}=x_{j}+iy_{j}$ denotes the position of the $j$th electron
as a complex number.  To see that $D$ attaches two
vortices to each electron, fix all $z_{j}$'s in $D$  except
$z_{1}$. As $z_{1}$ is taken around in a closed loop around any other
electron, the wave function gains a phase of $4\pi$, i.e., each
electron sees  two vortices on every other electron.
The resulting CF wave function is
\begin{equation}
\Phi^{CF}_{\nu^*,n,\alpha} = {\cal P}
D^{m}\;\Phi_{\nu^*,n,\alpha}\;\;.
\label{cfwf}
\end{equation}
${\cal P}$ projects the wave function onto the lowest LL of electrons
to obtain states appropriate for the large $B$ limit.
We will call the process of obtaining the CF state $\Phi^{CF}$ from a
given electron state $\Phi$ (which involves multiplication by $D$ and
then projection on to the lowest LL) ``composite fermion
transformation".

The CF theory asserts that the eigenstates of interacting electrons,
denoted by $\chi$, are
well described by the CF wave functions, i.e.,
\begin{equation}
\chi_{\nu,n,\alpha}=\Phi^{CF}_{\nu^*,n,\alpha}\;\;.
\label{cf2}
\end{equation}
The wave function of composite fermions
on the right hand side of Eq.~(\ref{cf2}) describes an electron system
at filling factor $\nu$, given by
\begin{equation}
\nu=\frac{\nu^*}{2m\nu^*+1}\;\;.
\label{ff}
\end{equation}
This equation can also be written in terms of a relationship between
the effective magnetic field experienced by the composite fermions,
$B^*=\rho\phi_{0}/\nu^*$, and the external field $B=\rho\phi_{0}/\nu$
as:
\begin{equation}
B^*=B - 2m \rho\phi_{o}\;\;.
\label{relB}
\end{equation}
There is another way of seing why the composite fermions experience a
smaller effective magnetic field. Take a composite fermion around a
loop enclosing area $A$. The total phase associated with this path is
\begin{equation}
2\pi( BA/\phi_{0}- 2m \rho A)
\end{equation}
where the first term is the usual Aharonov-Bohm phase and the second
term is the phase due to the vortices bound to other composite fermions
inside the loop. Equating this to $2\pi B^* A/\phi_{0}$,
the Aharonov-Bohm phase due to an
effective field $B^*$, produces Eq.~(\ref{relB}). This further clarifies
why binding of vortices to electrons leads to an effective
renormalization of the magnetic field.

The principal feature of the
composite fermion theory is that  the strongly correlated
liquid of interacting electrons at magnetic field $B$ (or filling
factor $\nu$) resembles a gas of non-interacting composite fermions at
magnetic field $B^*$ (or filling factor $\nu^*$). The composite
fermions form quasi-LL's \cite {quasi}, analogous to LL's of electrons.
There are strong
experimental reasons to believe this scenario. To begin with,
it provides a simple explanation of the origin of the FQHE \cite
{Tsui}.  IQHE of composite
fermions ($\nu^*=n$) corresponds to electron filling factors
$\nu=n/(2mn+1)$. These are precisely the observed fractions \cite {JG},
demonstrating that the FQHE of electrons is
simply the IQHE of composite fermions.
Many properties of the system in the FQHE regime can be understood in
terms of weakly interacting composite fermions:
the resistance peaks for high quality samples
occur half-way between two successive quasi-LL's
(i.e., at $\nu^*=n+1/2$) \cite {GJS}; the excitation gaps are successfully
modeled as quasi-cyclotron energies of composite fermions \cite {Du}; and the
oscillations in resistance at relatively high temperature are well described
in terms of Shubnikov-de Haas oscillations of composite fermions \cite
{Leadley}. There is good evidence that the composite fermion description is
also valid for the compressible state at $\nu=1/2$, where they see
vanishing effective field, and form a quasi-Fermi sea, investigated by
Halperin, Lee, and Read \cite {HLR}. The cyclotron orbit of composite fermions
for small $B^*$ has been observed in several recent experiments \cite
{Kang}.

There is also strong evidence for the validity of the
CF theory from numerical studies
on few electron systems \cite {Dev,Wu,others,Rezayi}.
While these numerical calculations are
limited to systems with few (6-10) electrons, they
have the virtue of being exact, and
thereby tell us in an unprejudiced manner how
much and how far the CF theory may be trusted.
In detailed numerical studies on few electron systems,
it has been found that there is a low-energy band in the
spectrum of interacting electrons at $B$, which
is remarkably similar to the low-energy band of non-interacting fermions
at $B^*$, establishing the formation of composite fermions over a wide
range of filling factors \cite {Dev}.
Furthermore, the CF wave functions, which involve no adjustable
parameters, have been found to be almost
identical to the exact numerical wave functions in this band.
For the special case of the ground state at $\nu=1/(2m+1)$, the CF wave
function is identical to the Laughlin wave function \cite {Laughlin},
which had already been known to be a good representation of
the exact ground state at $\nu=1/(2m+1)$
\cite {Fano,Haldane}. These studies provide a compelling
evidence for the validity of the CF theory for the low-energy physics.

This work investigates the excited states, which are also expected to
form bands from the
composite fermion theory. The principal motivation is experimental.
For an explanation of the FQHE, which is a low-temperature phenomenon,
an understanding of the states below the gap is sufficient, but
several other experiments require understanding of the
higher bands. Optical recombination experiments show evidence for
bands in the spectrum \cite {optical}. Raman experiments probe
the collective mode excitations of the FQHE system \cite {Pinczuk}.
Another situation when the higher bands become relevant
is when the gap disappears, which happens when the composite fermions
fill a large number of quasi-LL's, i.e., in the limit when $\nu=n/(2n+1)
\rightarrow 1/2$, when composite fermions form a Fermi sea.  Our study
allows us to make some qualitative statements about the nature of the
compressible state at $\nu=1/2$.

Another motivation for the present study is simply to explore
further the formal structure of the CF theory, and determine or
extend the limits of its validity.
A summary of our findings is as follows. We have studied two to three
bands for several systems. We find that the CF theory continues to
provide a good description of these bands. However, this happens only
as a result of many magical cancellations.  As will be discussed
below, the system of non-interacting electrons at $\nu^*$ contains
``too many states", but
just the right number of them are eliminated in the CF transformation, and
the remainder provides a good description of the spectrum of
interacting electrons at $\nu$.

{\bf II. THE SPHERICAL GEOMETRY}

Our numerical calculations employ the spherical geometry \cite {Haldane}
in which  $N$ electrons move on the surface of a sphere under
the influence of a radial magnetic field.
The flux through the surface of the sphere is $N_{\phi}hc/e$, where
$N_{\phi}$ is an integer. We consider spinless electrons confined to
their lowest LL, as appropriate in the limit of large magnetic
field. The total orbital angular momentum $L$ is analogous to the
wave vector of
the planar geometry; larger $L$ corresponds to larger wave vector
\cite {Haldane}.

We summarize here some basic results about the single electron wave
functions, and refer the interested
reader to the Appendix and two very useful papers by Wu and Yang \cite
{WuYang} for more detailed information.
The eigenstates in the spherical geometry are called ``monopole
harmonics" $Y_{S,l,m}$.  For $S=0$ they are the usual spherical
harmonics $Y_{l,m}$.  The quantity $l=|S|,\;|S|+1,\;...$ denotes the
orbital angular momentum, and $m=0,\;\pm 1,\;\pm 2,\;... \pm l$ is its
component in the z direction. (The notation $m$ is also used for the
amount of vortices attached to each electron; the meaning should be
clear from the context.) Note that the minimum value of $l$ is
$|S|$, and it can take either integer or half integer values.

The energy of an electron in the $l$th angular momentum shell is given
by
\begin{equation}
E_{l,m}=\hbar^2\frac{l(l+1)-S^2}{2m_{e}R^2}
\end{equation}
where $m_{e}$ is the electron band mass and $R$ is the radius of the sphere.
Using the fact that the flux through the surface of the sphere is $2S$
flux quanta, i.e., $$4\pi R^2 B= 2S\;(hc/e)\;\;,$$ the energy can be
rewritten as
\begin{equation}
E_{l,m}=\frac{l(l+1)-S^2}{2S}
\end{equation}
in units of the cyclotron energy $\hbar\omega_{c}=\hbar eB/m_{e}c$.
The  angular momentum shells of the spherical geometry are thus
analogous to the LL's of the planar geometry. The lowest ($n=1$)
LL is given by
$l=|S|$ shell, and the $n$th by the $l=|S|+n-1$ shell.
The degeneracy of the lowest ($n=1$) LL is $2|S|+1$
and increases by two for each successive LL.

We will assume large Zeeman energy throughout this work,
and will therefore work with spinless
electrons.  The Coulomb interaction
commutes with the total angular momentum ($L$), which will be used to
label the eigenstates. All eigenstates in a given $L$ multiplet
(i.e., states with different  $L_{z}$) have the same energy. Therefore,
it is sufficient to work in the subspace of smallest $L_{z}$, provided
it is remembered that each state in this subspace denotes a multiplet.
We will usually work in the subspace $L_{z}=0$; one state in
this subspace represents a multiplet of $2L+1$ degenerate states of
the full Hilbert space.

The Eqs.(1-5) are easily generalized to spherical
geometry \cite {Dev,Wu}: the Jastrow factor $D$
in the spherical geometry is the square of the
wave function of the lowest filled LL of spinless electrons, i.e.,
$D=\Phi_{1}^2$ (which is also true in the planer
geometry, aside from the exponential factors).
We also find it more convenient to label the wave
functions by $S$, rather than the filling factor $\nu$ or the magnetic
field $B$, so that the
Eqs.~(\ref{cfwf}) and (\ref{cf2}) are written as
(suppressing some of the subscripts)
\begin{equation}
\chi_{S}=\Phi^{CF}_{S^*}={\cal P} \Phi_{1}^{2m}
\Phi_{S^*}\;\;.
\label{cfwfsp}
\end{equation}
Since, for $N$ electrons a filled LL is obtained for $2S=N-1$, the
Eq.~(\ref{cfwfsp}) leads to the following relation between $S$ and $S^*$:
\begin{equation}
S=S^*+m(N-1)\;\;.
\label{relS}
\end{equation}
Clearly, this reduces to Eq.~(\ref{ff}) in the thermodynamic limit.
An important feature of the CF theory is that the CF transformation
conserves the total  angular momentum $L$. Some details of how we
obtain the projected CF wave functions are given in the appendix.

{\bf III. THE BAND STRUCTURE}

We start by noting that there  are three systems in this problem: (i)
interacting electrons at $S$ (`the
$\chi$ system'); (ii) non-interacting electrons at $S^*=S+m(N-1)$
(`the $\Phi$ system'); and (iii) non-interacting composite fermions at
$S^*$ (`the $\Phi^{CF}$ system'). We know everything about the
$\Phi$ system, and would like to learn as much as possible about
the $\chi$ system. As stated earlier, the $\Phi$ system contains
bands of degenerate states. These bands translate into bands of the
$\Phi^{CF}$ system, and eventually into bands of the $\chi$ system.
Such band structure was anticipated in Eqs.~(\ref{cfwf}) and (\ref{cf2})
in writing the
subscripts.  Besides predicting that there are bands in the $\chi$
system, which is already a non-trivial accomplishment (since all states are
degenerate in the absence of interactions), the composite fermion
theory claims that the states in each band of the $\chi$ system
have a one-to-one mapping with
the states in the corresponding
band of the composite fermion ($\Phi^{CF}$) system,
and the actual eigenstates are accurately given by
the CF wave functions in Eq.~(\ref{cfwf}) or (\ref{cfwfsp}).

An important feature of  the composite fermion approach is that
the Hilbert spaces of the $\Phi$ and $\chi$ systems are of
drastically different sizes. This
is most easily seen for a finite number of
electrons: the Hilbert space of the $\Phi$ system is
infinite, since all LL's are allowed, while that of the $\chi$ system
is finite, since only the lowest LL is allowed. Therefore, {\em it is not
possible, even in principle,
for all bands of the $\Phi$ system to be analogous to the
bands of the $\chi$ system}. On the other hand, by construction,
the Hilbert space of the $\Phi^{CF}$ system is identical to that of
the $\chi$ system.  Therefore, the analogy between the $\chi$ and the
$\Phi^{CF}$ systems can be expected to be more general.
It was found earlier \cite {Dev}, that the lowest
bands of the three systems are indeed analogous in a wide range of
filling factors: i.e., they have the same number of states,
with identical quantum numbers.
This was a remarkable result, showing that the low-energy
dynamics of interacting
electrons at $S$ is similar to that of non-interacting electrons at
$S^*$. In the present work, we extend this result to higher bands.

We have studied four systems with $(N,S)=$ (6,7.5), (7,9), (8,8.5),
(6,8). The first two represent the state at $\nu=1/3$, the third
lies somewhere between 1/3 and 2/5, and the fourth has $\nu>1/3$.
The exact (Coulomb) low-energy spectra
of these  systems are shown in Figs. 1-4 (a). There is a clearly defined
lowest band. The second bands can also be identified rather easily. The
third and higher bands are broad, mix with other bands, and are
only poorly defined.

The spectra of the corresponding non-interacting electrons
(the $\Phi$ system) at $(N,S^*)=$ (6,2.5), (7,3), (8,1.5), and (6,3)
are also shown in Figs. 1-4 (c). There are bands of degenerate
states separated by the cyclotron energy; the number above each line
shows the number of orthogonal multiplets at that energy.
The second band is obtained by exciting one electron by one LL, and
the states in the third band contain either one electron excited by two
LL's or two electrons by one LL.

The largest $L$ in each band of the $\chi$ system is correctly
identified by the $\Phi$ system. The lowest bands of the two systems
also match perfectly. For the 1/3 state, the lowest band
contains only one state,
which is the one filled LL state in the $\Phi$ system. A CF
transformation of this state produces the Laughlin state, which is
known to be quite close to the ground state of the 1/3 state.
However, the higher bands of the $\chi$ and the $\Phi$ systems do not match.
(In Figs. 3 and 4, the higher bands of the $\chi$ systems are
reasonably well defined, and a lack of matching is quite clear.
In Figs. 1 and 2, the third band is not very
well defined. However, if we construct bands of the $\chi$ system
in analogy with the
$\Phi$ system, they would contain very high energy states at some low
values of $L$, while some low-energy states at larger values of $L$
will be outside the bands.)
The second bands in Figs. 1(c) and 2(c) have one
multiplet each at $L=1,2, ... N$, but,  the corresponding bands of the
$\chi$ system in Figs. 1(a) and 2(a) do not have any multiplet at $L=1$.
The second bands of Figs. 3(a) and 4(a) have
even less similarity to the second bands of 3(c) and 4(c).
This leads us to conclude that the exact correspondence of the
$\Phi$ and the $\chi$ systems is limited to the lowest band.

Next we consider the CF system ($\Phi^{CF}$)
at $S^*$. We construct the CF states by a CF
transformation of the states of the $\Phi$ system.
Several surprising things happen.

(i) The
$L=1$ state of the second bands of Figs. 1(c) and 2(c) are annihilated upon
projection (i.e., their projection is identically zero \cite{F10}); as a
result, the corresponding $\Phi^{CF}$
systems do not have an $L=1$ state in the second bands \cite {Dev}.

(ii) Orthogonal states of  the $\Phi$ system do not,
in general, produce orthogonal states upon the CF transformation.
(Of course, states with different $L$ produce orthogonal states.)

(iii) The states obtained from the second band of Figs. 3(c) and 4(c)
are not even linearly independent in general.
We construct an orthogonal basis
using the Gramm-Schmid procedure, which is
smaller than the ($\Phi$) basis we started with.

(iv) Consider states produced in this way at a given $L$. They
contain {\em all} states of the lower bands at that $L$, i.e.,
the states in the lower band can be exactly
expressed as a linear combination of the states thus produced.
We construct a still smaller basis, which is
orthogonal to the lower
band CF states, and identify it with the second band of the CF system.

(v) The situation for the third bands of Figs. 1(c) and 2(c) is similar.
First, the number of linearly independent states is smaller after the
CF transformation. Second, the lower band states are contained in the
CF states produced from the third band of electrons. We again construct
a basis which is also orthogonal to the lower band states, and
identify it with the third band of the corresponding CF system.

(vi) The number of orthogonal multiplets at various $L$ in the higher
bands of the $\Phi^{CF}$ systems is shown in Figs.1-4 (b).
It is rather remarkable that the bands of the $\Phi^{CF}$ system
exhibit a perfect one-to-one correspondence with
the bands of the $\chi$ system.

(vii) In order to provide a microscopic confirmation of the CF theory,
we consider a sub-basis of CF states at a given $L$ in a given band.
We wish to compare it with the corresponding sub-basis of $\chi$ states.
According to the CF theory, we need to diagonalize the Coulomb
Hamiltonian in the CF sub-basis, determine the eigenstates, and calculate
the overlap of each CF eigenstate with the corresponding
eigenstate of the $\chi$ system. We define ``generalized overlap" as the
geometric mean of all these overlaps, given by $[\prod_{j}O_{j}]^{1/J}$,
where $O_{j}$ is the overlap of the $j$th states of the two sub-bases, and
$J$ is the number of states in the CF (or the $\chi$) sub-basis.
A large generalized overlap indicates that the CF sub-basis is close to
the $\chi$ sub-basis. Other than the fact that the generalized overlap
gives a single measure of the similarity between two bases, it also
has the advantage that its calculation does not require any
diagonalization. It is given by
\begin{equation}
[Det <\Phi_{j}^{CF}|\chi_{j}>]^{1/J}\;,
\end{equation}
where the states $\Phi_{j}$ and $\chi_{j}$ are the properly orthonormalized
states of the two sub-bases.  The (generalized) overlaps are shown on
the figures themselves (See Figs. 1b-4b). They are generally
close to unity, indicating the microscopic validity of
the CF description for higher bands.

The above discussion gives rules for constructing the bases for
successive bands of interacting electrons using the CF framework.
For all four systems studied above, the bases constructed in this
manner provide a systematic description of the
energy spectra of interacting electrons. We believe that these rules
are of quite general validity in the range of filling factors where
the composite fermion description is applicable for the lowest band.

We would like to emphasize that there is no {\em a priori}
reason for expecting
cancellations in going from the electron system ($\Phi$) to the
CF system ($\Phi^{CF}$). Of course, since the Hilbert space of the
latter is smaller than that of the former, eventually, for much higher
bands, such cancellations are inevitable. However, the lowest two or
three bands of the systems studied above cover only a small part of
the full Hilbert space of the $\Phi^{CF}$ system. Thus, it is rather
non-trivial that {\em seemingly} distinct  trial wave functions, constructed
from {\em orthogonal} $\Phi$  states, quite frequently turn out to
be {\em mathematically} linearly dependent. This shows that the CF
trial wave functions possess certain
non-trivial mathematical symmetries. A formal appreciation of
these symmetries will require a better
understanding of the projection operator. In order to describe the
thermodynamics of the FQHE state, it would be quite useful to
have some general understanding of how many states are eliminated
in going from the $\Phi$ system to the $\Phi^{CF}$ system.

It should also be clear that the
reduction in the number of states in the CF transformation
is not a finite size effect. Even though our systems are small,
the number of states in the first
three bands is only  a small fraction of the total number of states.

{\bf IV. COLLECTIVE MODES OF COMPOSITE FERMIONS}

For incompressible states, the second band contains precisely one
multiplet at each $L$ in a certain range of $L$ (see, e.g., Figs. 1
and 2). This will be called the
collective mode  branch (also known as the exciton or the plasmon
branch). The angular momentum $L$ is
roughly proportional to the momentum $q$ of the planar geometry \cite
{Haldane}.

In the usual three dimensional electron gas systems, when the
density operator ($\rho_{q}$) acts on the ground state, it creates
(for small $q$) a state
whose oscillator strength is sharply peaked at some energy.
It is called, for obvious reason, the ``collective mode". A single
mode approximation (SMA) assumes that the peak is a delta function.
Soon after Laughlin's theory of the ground state at $\nu=1/(2m+1)$,
Girvin, MacDonald and Platzman (GMP) \cite {GMP} developed an
SMA theory for the collective modes of the Laughlin states.
Straightforwardly extending it to general fractions,
the GMP wave function for the collective mode of the
$\nu=n/(2n+1)$ state is constructed by applying the projected electron
density operator to the ground state as
\begin{equation}
{\cal P}\rho^{L}\Phi_{n}^{CF}={\cal P}\rho^{L}[{\cal P}D\Phi_{n}]\;\;,
\label{elcm}
\end{equation}
where $\Phi_{n}^{CF}={\cal P}D\Phi_{n}$ is the ground state at
$\nu=n/(2n+1)$, and
\begin{equation}
\rho^{L}=\sum_{i=1}^{N}Y_{L,0}(\Omega_{i})
\end{equation}
is the density operator at $L$.
It has been shown that this wave function
provides a good description of the low- and intermediate-$L$
states at $\nu=1/3$, as shown in Table I.  It, however,
fails at large $L$. It also does not work as satisfactorily at other
fractions \cite {Su,Reynolds}.

The low-energy excitations of the FQHE state
are described in terms of composite fermions, rather than electrons.
Therefore, it might seem reasonable to consider collective modes of
{\em composite fermions}. (Since an excited composite fermion is a
quasielectron, this could also be called a collective mode of
quasielectrons. Similar suggestions have been made in the
past. See, for example \cite {Laughlin84}.) In contrast to
Eq.~(\ref{elcm}), which gives the collective mode of electrons,
the collective mode of composite fermions of the $\nu=n/(2n+1)$ state
is constructed by applying the CF transformation on the collective
mode state at $\nu^*=n$ \cite {Kallin}.  Its wave function is given by
\begin{equation}
{\cal P}D\rho^{L}(n\rightarrow n+1)\Phi_{n}\;\;,
\label{cfcm}
\end{equation}
where $\rho^{L}(n\rightarrow n+1)$ creates the ``fundamental" collective
mode at $\nu^*=n$, which involves excitation of  one electron from
the $n$th LL to the $(n+1)^{st}$ LL.
For the state at $\nu^*=1$, the excited electron in the second LL has
angular momentum $(N+1)/2$ whereas the hole left behind in the lowest LL has
angular momentum $(N-1)/2$, so the allowed values of the total angular
momentum are $L=1,\;
2,\;... \; N$. For the particle-hole excitation at $\nu^*=2$, the
excited electron in
the third LL has angular momentum $N/4$ whereas the hole in the second
LL has angular momentum $(N+4)/4$, implying that the allowed values of the
total angular momentum are $L=1,\;...,(N+2)/2$. The wave function of
the collective mode state is uniquely determined by symmetry for
any given $L$, with no fitting parameter. (In particular, it does not
depend on the cyclotron energy.) Upon CF transformation, the
wave function of the collective mode of composite fermions is obtained
for these values of $L$, again with no adjustable parameter.

First we note that the collective mode branch in the exact
diagonalization study has maximum $L$ consistent with the prediction
of the CF theory. The range of
the collective mode branch is predicted correctly by the
CF theory for 3/7 state as well; for spectra, see Refs.
\cite {Reynolds,Simon}.  The electron (i.e., GMP) collective mode,
on the other hand, is defined upto much higher values of $L$.
Table I shows that Eq.~(\ref{cfcm})  provides a good approximation of
the entire collective mode branch for both 1/3 and 2/5.
A good overlap guarantees that the energy of the CF state
is very close to the exact energy; the Coulomb energies of the CF
collective mode states are shown on the Figures themselves, and agree
well with the exact energies.
There is little doubt that the collective mode branches of other
incompressible states can also be modeled successfully
as the collective mode of composite fermions.

Eq.~(\ref{cfcm}) gives the wave function of the
collective mode in which one composite fermion is excited from the
topmost filled quasi-LL to the lowest empty quasi-LL.
One may wonder if there are more collective modes of composite
fermions, corresponding to excitations of composite fermions across
two or more quasi-LL's. Analogous modes indeed exist for electrons
in the IQHE regime \cite {Kallin}. In order to investigate this
question for $\nu=1/3$, we construct states
${\cal P}D\rho^{L}(1\rightarrow 2)\Phi_{1}$, ${\cal
P}D\rho^{L}(1\rightarrow 3)\Phi_{1}$, and ${\cal
P}D\rho^{L}(1\rightarrow 4)\Phi_{1}$.
We find that for $L=$ 2 and 3, these are {\em mathematically}
identical for systems with $N= $ 4, 5, 6, and 7, and presumably for
arbitrary number of electrons.  (Note that there
is no $L=2$ state for the $1\rightarrow 4$ excitation.)
The total number of states is fairly large for
$L=2$ and 3, and therefore the fact that
the various composite particle-hole states are
identical is rather non-trivial. For $L=4$ and 5, the various
composite fermion modes are not identical, but they still have a
reasonably good overlap with the $1\rightarrow 2$ state.
These results suggest that there is only one collective mode of
composite fermions, despite the fact that there are several
in the electron system $\Phi$. This is a striking example of the
breakdown of a one-to-one correspondence in the higher bands of $\Phi$
and $\Phi^{CF}$.

It might be argued that our conclusion regarding the number of
collective modes might not remain true in the presence
of a small amount of LL mixing, always present in experiment. In this
case, it would seem natural {\em not} to project the composite fermion
states on to the lowest LL, so they would not be identical.
However, clearly, the unprojected composite fermion states
will simply provide {\em different approximations} for the {\em same}
collective mode. The essential point, quite obvious on physical
grounds, is that the {\em intra-LL} spectrum of interacting
electrons, or the {\em number} of {\em intra}-LL collective modes,
cannot change in any essential manner when a small amount of LL
mixing is allowed. Therefore, our conclusion, obtained with the
lowest LL approximation, should remain valid for the more
realistic situation of a large though not infinite $B$.

At small $L$, the CF description of the
collective mode branch of the 1/3 state
is not as good as at higher $L$. The relatively
poor performance of the CF theory may be attributed
to the fact that the collective mode branch is coming close to the
higher bands; the exact state mixes with higher band states to reduce
its energy, whereas any such mixing
is neglected in the CF approach. For 2/5, the description becomes
worst at {\em intermediate} values of $L$, where the collective mode
is the closest to the higher bands.

It is intuitively reasonable that, for small
$L$, the collective mode can be thought of either as a mode of
electrons or as a mode of composite fermions, since small
(relative) displacements of composite fermions are indistinguishable
from small displacements of electrons.
This is supported by the fact that both the CF and the GMP
theories provide reasonable descriptions of the true collective mode for
small $L$, both for 1/3 and 2/5. (For the 1/3 state, the GMP theory
does slightly better than the CF theory for small $L$. Also note that
the $L=1$ state is annihilated by the projection operator in both
schemes. We note that the work of Reynolds and d'Ambrumenil \cite
{Reynolds}
also finds that the SMA works quite well at small $q$ for 2/5 and
3/7.) This suggests that there is one and only one well-defined
{\em electron} collective mode for all incompressible states in the limit of
small $q$.

Raman experiments observe collective mode at small wave vectors, where
it may be interpreted as either an electron or a composite fermion
collective mode.
Pinczuk {\em et al.} \cite {Pinczuk} have reported observation of the
collective mode of the 1/3 state in Raman experiments. No
collective modes have yet been observed at other fractions.
Recently \cite {Pinczuk1}, a broad secondary peak has also been observed
at finite wave vectors at $\nu=1/3$; we believe that this is
an observation of the third band of composite fermions.
Chen and Quinn \cite {Chen} have suggested that the appearance of
double peaks in photoluminescence experiments \cite {optical}
is also due to the formation of bands, explained naturally within the CF
framework.

{\bf V. COMPARISON WITH CHERN-SIMONS THEORY}

Our few electron calculations have
shown that the CF-wave-function scheme is qualitatively and
quantitatively trustworthy, not
only for the low-energy physics but also for the excitations.
With further improvements in techniques for handling the projection
operator for bigger systems, the CF-wave-function scheme should provide good
quantitative approximations for various quantities in the
thermodynamic limit.

The CF wave functions were motivated by a physical picture in which
an even number of flux quanta are attached to each electron of $\Phi$
and then the flux is smeared to obtain electrons at a new magnetic
field \cite {Jain}. A Chern-Simons (CS)
field-theoretical scheme has also been developed based on this
picture \cite {Lopez,HLR}. In this scheme,
the $\chi$ system of interacting electrons at $\nu$
is mapped on to the $\Phi$ system of non-interacting electrons at
$\nu^*$ at the mean-field level.  Fluctuations about the mean-field state are
then treated perturbatively, usually
at the random-phase-approximation (RPA) level \cite {Lopez,HLR}.
In spite of the fact that both the wave-function and CS approaches
are guided by the same physics,
the precise relationship between them is not known.
It has been asserted \cite {Lopez} that RPA is equivalent to multiplication
by the Jastrow factor, but no proof exists.

The main implication of our results for the CF-CS scheme is
that understanding of {\em excitations} in this scheme
is likely to be a rather subtle
and complicated issue. As stated above, the
CF-CS theory proposes to approach the $\chi$ system perturbatively
starting from the $\Phi$ system.  However,
the $\Phi$ system has the `correct' number of states only  in
the lowest band; many states in the higher bands of this
system are `unphysical' in the sense that they
do not have any analog in the $\chi$ system (some
are annihilated upon the CF transformation, while some produce
states of the {\em lower} bands).
It is not understood at present how perturbation theory starting from the
$\Phi$ system will get rid of the spurious unphysical states and avoid
or overcounting.

This is illustrated in the investigation of the
collective modes using the Chern-Simons RPA approach \cite
{Lopez2,Simon}. It finds a {\em series} of
{\em electron} collective modes for all incompressible states, {\em
including the Laughlin states}. These are analogous to the electron
collective modes at integer fillings.  Another feature of the
CF-CS approach is that the ``fundamental" inter-LL mode of the $\nu^*=n$
state (i.e., the mode with energy $\hbar\omega_{c}$, which
corresponds to $n\rightarrow n+1$
electron-hole excitation) does not produce any mode at $\nu=n/(2n+1)$;
it is pushed up to the cyclotron energy \cite {Lopez}.
This is actually crucial in recovering the Kohn's theorem at $\nu$.
Thus, the lowest energy electron collective mode at $\nu$ is expected
to correspond to the  $2\hbar\omega_{c}$ collective mode of the
$\nu^*=n$ state. (Lee and Wen \cite {LeeWen} have also made a similar
suggestion in order to explain the absence of the $L=1$ state of the
collective mode.)
These results seem to be inconsistent with our study, where several
collective modes of the IQHE state are found to map into a single
collective mode of the FQHE state (at least for small wave vectors),
and the entire collective mode of the FQHE state is
well described as the {\em fundamental} mode of composite fermions.

Our study also has consequences for the nature of the compressible
state at $\nu=1/2$. To understand this, let us approach the 1/2 state
as the $n\rightarrow \infty$ limit of the
$n/(2n+1)$ sequence of incompressible FQHE states.
Let us assume that the CF description continues to remain valid for
arbitrarily large $n$ (which is unfortunately not testable in
finite system studies). Then, one would expect the
ground state to correspond to $n$ filled LL's of electrons, but
the excitations do not have a one-to-one correspondence with
the excitations of the non-interacting electron state at $\nu^*=n$.
In the FQHE regime, these excitations are separated by a gap, and do
not have much influence on the low-energy physics of the problem.
However, the gap vanishes in the limit $n\rightarrow\infty$,
as the sequence approaches $\nu=1/2$.  Two comments
pertaining to this limit may be made in light of the results obtained
above.  (i) Since infinite filled Landau levels is nothing but a Fermi
sea, the {\em ground state} of interacting electrons at $\nu=1/2$
is likely to resemble a Fermi sea of composite fermions with a well defined
Fermi surface, as proposed by Halperin, Lee, and Read \cite {HLR}. Recent
experiments lend strong support to the existence of a Fermi sea at
$\nu=1/2$ \cite {Kang}.  (ii) However, the {\em excitations} of the
1/2 system do {\em not} have a one-to-one correspondence
with the excitations of non-interacting {\em electrons} at zero
magnetic field. Therefore, despite the existence of the Fermi
surface, the $\nu=1/2$ state is not a conventional Landau Fermi
liquid in the sense of state counting.

This conclusion may seem inconsistent with that of
Rezayi and Read \cite {Rezayi}, who recently attempted an investigation
of  the 1/2 state by finite size numerical study. They considered
electron systems at $S=N-1$, which map onto CF systems at zero
effective flux ($S^*=0$), and  found that the lowest
band is well described in terms of the CF theory, in
agreement with earlier results \cite {Dev}. This was interpreted as a
confirmation of the idea that the 1/2 state is a Fermi liquid.
However, this study only deals with an {\em `atom'} of composite fermions
at zero flux, which is, of course, not a Fermi liquid.  (Here the `atom'
has the special feature that electrons are constrained to
move on a spherical surface.) Approaching a
Fermi liquid from the atomic limit is rather subtle. To see this,
take the limit $N\rightarrow\infty$ while keeping the density fixed.
The kinetic energy gap between successive angular momentum shells vanishes as
$1/\sqrt{N}$ and the band structure is completely destroyed in this
limit. Let us consider the situation when the kinetic energy gap is
very small (say, compared to the interaction band-width)
but non-zero. The low-energy excitations now consist of not only the
excitations {\em within} the lowest angular momentum shell, but also
excitations {\em across} several shells. The crucial point is that
{\em all} these excitations are
needed to produce a conventional Fermi liquid. Therefore, in
order to prove the existence of a conventional Fermi liquid
at $\nu=1/2$, it is necessary to establish a one-to-one correspondence
between all such (low-energy) excitations, which, as shown in the
present work, does not exist \cite {Jain2}.

While our study cautions against too literal an interpretation of
the 1/2 state as a Fermi liquid, it does show that it is
{\em at least related}
to a Fermi liquid, and has a ground state that is likely a filled
Fermi sea. Therefore, it is quite possible that eventually a
Fermi-liquid-like description with renormalized parameters will prove
sufficient.  We believe that the origin of the qualitatively different
predictions of the CF-wave-function and the CF-CS approaches lies in
the fact that no natural way is known of incorporating
the large $B$ limit, or of implementing the lowest LL constraint in
CS approach. This is indicated by the fact that the final
formulas of the CS theory explicitly contain the band mass of the
electron, which is not a parameter in a lowest LL theory.

{\bf VI. CONCLUSION AND ACKNOWLEDGEMENTS}

We have investigated the excitation spectrum of interacting electrons
in the FQHE regime. We find that
the explanation of the excited bands is subtle, but
in principle possible within the composite-fermion wave-function
approach. The Hilbert space of non-interacting electrons at $\nu^*$
is much bigger than the relevant (i.e., lowest-LL) Hilbert space
of interacting electrons at $\nu = \nu^*/(2\nu^*+1)$, but
the CF transformation eliminates just the right number of states
from each band to bring the theory in agreement with the
actual band structure of the interacting electrons at $\nu$.

We are grateful to A.H. MacDonald, and A. Pinczuk for useful
conversations. This work was supported in part by the NSF under
Grant No. DMR93-18739. A short report of parts of this work has been
submitted elsewhere.

{\bf APPENDIX}

{\bf Coulomb interaction}

We will not need the explicit form of the monopole harmonics.
Their following properties will be sufficient for our purposes.
\begin{equation}
Y_{S,l,m}^*=(-1)^{S+m}\;Y_{-S,l,-m}
\end{equation}
\begin{eqnarray}
Y_{Slm}Y_{S'l'm'}& = & \sum_{l''}(-1)^{l+l'+l''+2(S''+m'')}
\left(\frac{(2l+1)(2l'+1)}{4\pi (2l''+1)}\right)^{1/2}
\nonumber \\
&& <lm,l'm'|l''m''><lS,l'S'|l''S''>\;Y_{S''l''m''}
\end{eqnarray}
where $S''=S+S'$ and $m''=m+m'$. Thus, the product of two spherical
harmonics at $S$ and $S'$ produces a linear combination of
spherical harmonics at $S+S'$.
\begin{eqnarray}
\int
d\Omega\;Y_{S_{1}l_{1}m_{1}}\;Y_{S_{2}l_{2}m_{2}}\;Y_{S_{3}l_{3}m_{3}}
&  =   (-1)^{l_{1}+l_{2}+l_{3}-m_{3}-S_{3}}
\left(\frac{(2l_{1}+1)\;(2l_{2}+1)}{4\pi(2l_{3}+1)}\right)^{1/2}
\nonumber \\
&
<l_{1}\;m_{1},l_{2}\;m_{2}|l_{3}\;-m_{3}>\;
<l_{1}\;S_{1},l_{2}\;S_{2}|l_{3}\;-S_{3}>
\end{eqnarray}
provided $S_{1}+S_{2}+S_{3}=0=m_{1}+m_{2}+m_{3}$.
Here, $d\Omega$ is the integral over the angular variables.

The coulomb matrix elements
can be easily evaluated using the properties of the monopole
harmonics.
The two-body matrix element is given by
\begin{eqnarray}
<l_{1}\;m_{1},l_{2}\;m_{2}| V(|{\bf
r_{1}}-{\bf r_{2}}|) | l_{1}^{'}\;m_{1}^{'},l_{2}^{'}\;m_{2}^{'}>=
\nonumber\\
\int d\Omega_{1} d\Omega_{2} Y_{Sl_{1}m_{1}}^*({\bf r_{1}})
Y_{Sl_{2}m_{2}}^*({\bf r_{2}}) \frac{e^2}{|{\bf r_{1}-r_{2}}|}
Y_{Sl_{2}'m_{2}'}^*({\bf r_{2}})
Y_{Sl_{1}'m_{1}'}^*({\bf r_{1}})\;\;.
\end{eqnarray}
The chord-distance between two electrons on sphere can be expressed
as:
\begin{equation}
\frac{1}{|{\bf r_{1}} - {\bf r_{2}}|} = \frac{4\pi}{R}\;
\sum_{l=0}^{\infty}\; \sum_{m=-l}^{l}
Y_{0,l,m}^{*}(\theta_{1},\phi_{1})
Y_{0,l,m}(\theta_{2},\phi_{2})\;.
\end{equation}
With this substitution, the two integrals decouple and can be
evaluated straightforwardly.

{\bf Projection}

The projected wave functions can be expressed as
\begin{equation}
{\cal P}\Phi_{1}^2\Phi_{S^*}=\sum c[\{M_{j}\}]\;|M_{1},...M_{N}>
\end{equation}
where $\Phi_{1}$ is the Slater determinant
wave function of the lowest filled LL at
$S_{1}=(N-1)/2$, $\Phi_{S^*}$ is the Slater determinant wave function at
$S^*$, and the $S$ of the product state is given by $S=2S_{1}+S^*$.
$M_{j}$ label z-components of
the single particle angular momentum $L=S$, restricted to the
range $-S\leq M \leq S$; they satisfy the restrictions
$M_{1}<M_{2}<...<M_{N}$ and $\sum_{j}M_{j}=0$, the latter because we
restrict to the $L_{z}=0$ subspace.
Our objective is to determine the coefficients $c[\{M_{j}\}]$.
The basic ingredient is the
formula for projection of the product of two single particle
eigenstates:
\begin{eqnarray}
{\cal P}Y_{Slm}Y_{S'l'm'}& = & (-1)^{l+l'+3S''+2m''}
\left(\frac{(2l+1)(2l'+1)}{4\pi (2S''+1)}\right)^{1/2}
\nonumber \\
&& <lm,l'm'|S''m''><lS,l'S'|S''S''>\;Y_{S''S''m''}
\end{eqnarray}
with $S''=S+S'$ and $m''=m+m'$.
In principle, the product $\Phi_{1}^2\Phi_{S^*}$ can be fully expanded, and its
projection can be obtained by repeated application of the above
equation.  We find it efficient first to expand $\Phi_{1}^2$ in terms
of basis states $|m_{1},...m_{N}>$ at $2S_{1}$, and
store the coefficients. (Note that since $\Phi_{1}^2$ is a boson wave
function, $m$'s do not have to be all distinct, as is the case with
$M$'s.) In order to obtain the coefficient
of a given basis state $|M_{1},...M_{N}>$, we go over each term
in the expansion of the Slater determinant
$\Phi_{S^*}$ and search for those values of $\{m_{j}\}$
which will provide the correct $\{M_{j}\}$. Then it is a matter of
complicated book-keeping to obtain the coefficient of
$|M_{1},...M_{N}>$. In general, the eigenstates of $L$ in the $\Phi$ system
are linear combinations of many Slater determinants;
to obtain the projection of this state,
we compute the projection for each Slater
determinant as above and add the projected states with the appropriate
coefficients. For further details, we refer the reader to Ref. \cite
{Wuth}.

{\bf Figure Captions}

Fig.1 (a) Low-energy part of the eigenspectrum of interacting electrons
in the lowest LL for the system $(N,S)=(6,7.5)$. The energies are in
units of $e^2/\epsilon l_{0}$ where $\epsilon$ is the dielectric
constant of the background material and $l_{0}$ is the magnetic
length. The horizontal lines show the bands. The dots show the energies
of the composite fermion states. (c) The low-energy
spectrum of the corresponding electron system at $(N,S^*)=(6,2.5)$.
The energy separation between the bands is equal to the cyclotron
energy. The integer near each dash shows the number of degenerate
(orthogonal) multiplets at that energy, with $2L+1$ states in each
multiplet.
(b) The low-energy spectrum of the CF system at $(N,S^*)=(6,2.5)$,
obtained from the system in (c) as explained in the text.
The separation between bands is the effective cyclotron energy of
composite fermions.
The integer above each dash shows the number of independent
multiplets, and the number below each dash shows the (generalized)
overlap of the CF basis with the exact eigenstates of (a).

Fig. 2 Same as Fig.1 for $(N,S)=(7,9)$ and $(N,S^*)=(7,3)$.

Fig. 3 Same as Fig.1 for $(N,S)=(8,8.5)$ and $(N,S^*)=(8,1.5)$.

Fig. 4 Same as Fig.1 for $(N,S)=(6,8)$ and $(N,S^*)=(6,3)$.

{\bf Table Captions}

Table I. The overlap of the electron and the CF
collective mode states with the exact collective mode states.
(a) 1/3 state with $N=6$. (b) 1/3 state with $N=7$. (c) 2/5 state with
$N=6$. (d) 2/5 state with $N=8$.  Some of these results have been
published before in some form, and are given here for completeness.
The results for the electron (GMP) collective mode have been given (in
slightly different forms) in Refs. \cite {Haldane,Su,Reynolds}. The results
for the CF collective mode (except for table b) have been published in
\cite {Dev}.

\begin{center}

\begin{tabular}{|c|c|c|c|c|c|} \hline
(a)$\;L$ & 2 & 3 & 4 & 5 & 6 \\ \hline
CF & 0.9484 & 0.9923 & 0.9915 & 0.9977 & 0.9800 \\ \hline
electron & 0.9796 & 0.9947 & 0.9818 & 0.8962 & 0.5435 \\ \hline
\end{tabular}

\vspace{5mm}

\begin{tabular}{|c|c|c|c|c|c|c|} \hline
(b)$\;L$ & 2 & 3 & 4 & 5 & 6 & 7 \\ \hline
CF & 0.9032 & 0.9793 & 0.9970 & 0.9906 & 0.9822 & 0.9853 \\ \hline
electron & 0.9313 & 0.9860 & 0.9903 & 0.9467 & 0.7301 & 0.3984 \\ \hline
\end{tabular}

\vspace{5mm}

\begin{tabular}{|c|c|c|c|} \hline
(c)$\;L$ & 2 & 3 & 4   \\ \hline
CF & 0.9999 & 0.9951 & 0.9983  \\ \hline
electron & 0.9995 & 0 & 0.9232  \\ \hline
\end{tabular}

\vspace{5mm}

\begin{tabular}{|c|c|c|c|c|} \hline
(d)$\;L$ & 2 & 3 & 4 & 5 \\ \hline
CF & 0.9946 & 0.9741 & 0.9818 & 0.9971  \\ \hline
electron & 0.9941 & 0.6604 & 0.7296 & 0.8927  \\ \hline
\end{tabular}

\end{center}


\begin{thebibliography}{99}
\bibitem{F10} The phrases `identically zero', `exact', or
`mathematically identical' refer to numerical results correct to
machine accuracy (i.e., to 15-16 significant figures). There is little
doubt that these results are exact.
\end{thebibliography}
\end{document}